\documentclass{article}
\usepackage[utf8]{inputenc}
\usepackage[margin=1in]{geometry}
\usepackage{graphicx}      
\usepackage{hyperref}
\usepackage{url}
\usepackage{amssymb}
\usepackage{amsmath}
\usepackage{bigints}
\usepackage{dsfont}
\usepackage{algorithm}
\usepackage{algpseudocode}
\usepackage{color}

\newtheorem{thrm}{Theorem}

\newtheorem{assump}{Assumption}

\def \R {\mathbb{R}}
\def \C {\mathbb{C}}

\def \diff {\mathrm{d}}
\def \statedim {p}
\def \controldim {q}
\def \noisedim {d}

\def \Qmat {Q}
\def \Rmat {R}
\def \truth {A_\star,B_\star}

\def \stabradii {c}

\def \stablength {\boldsymbol{\tau}}
\def \noofepochs {n}
\def \FBscale {\sigma_{{L}}}
\def \ditherstep {\epsilon}

\newcommand{\Mnorm}[2]{{\left\vert\kern-0.30ex\left\vert #1 
		\right\vert\kern-0.30ex\right\vert}}
\newcommand{\norm}[2]{{\left\vert\kern-0.30ex\left\vert #1 
		\right\vert\kern-0.30ex\right\vert}}

\newcommand{\eigmin}[1]{\boldsymbol{\lambda}_{\min} \left( #1 \right)}

\newcommand{\PP}[1]{%
	\mathbb{P}{\ifthenelse{ \equal{#1}{} }{}{\left(#1\right)}}
}%
\newcommand{\E}[1]{\mathbb{E}\left[#1\right]}

\newcommand{\RiccSol}[1]{\boldsymbol{{K}}\left(#1\right)}
\newcommand{\Optgain}[1]{\boldsymbol{{L}}\left(#1\right)}
\newcommand{\Gainmat}[1]{{L}_{#1}}

\newcommand{\optavecost}[1]{\overline{\mathcal{J}}^\star }

\newcommand{\optdisccost}[1]{\mathcal{J}_\gamma^\star}

\newcommand{\order}[1]{ \mathcal{O} \left(#1\right)}

\newcommand{\Amat}[1]{A_{\star}}
\newcommand{\Bmat}[1]{B_{\star}}
\newcommand{\CLmat}[1]{D_{#1}}
\newcommand{\estpara}[1]{\left[{A}_{#1},{B}_{#1}\right]^\top}
\newcommand{\estA}[1]{{A}_{#1}}
\newcommand{\estB}[1]{{B}_{#1}}
\newcommand{\estD}[1]{{D}_{#1}}
\newcommand{\MatOpAve}[2]{\Phi_{#1}\left(#2\right)}

\newcommand{\state}[1]{X_{#1}}

\newcommand{\statetwo}[1]{Y_{#1}}
\newcommand{\action}[1]{U_{#1}}
\newcommand{\itointeg}[4]{\int\limits_{#1}^{#2} {#3} \diff {#4}}
\newcommand{\indicator}[1]{\mathds{1}_{\left\{#1\right\}}}
\newcommand{\cov}[1]{\mathrm{Cov}\left(#1\right)}

\newcommand*{\BM}[1]{
	\mathbb{W}_{\ifthenelse{ \equal{#1}{} }{}{#1}}
}%
\newcommand{\dither}[1]{V_{#1}}
\newcommand{\BMcoeff}[1]{C_{\star}}
\newcommand{\dithercoeff}[1]{\sigma_{V}}

\newcommand{\normaldist}[2]{\mathcal{N} \left( #1, #2 \right)}

\newcommand{\mosteig}[1]{\boldsymbol{\overline{\lambda}} \left( #1 \right)}

\newcommand{\erterm}[1]{\Delta_{#1}}

\newcommand{\trans}[1]{D}
\newcommand{\mult}[1]{p_{#1}}

\newcommand{\posterior}[1]{\mathcal{D}_{#1}}

\newcommand{\empmean}[1]{\boldsymbol{\mu}_{#1}}
\newcommand{\empiricalcovmat}[1]{\boldsymbol{\Sigma}_{#1}}
\newcommand{\iidGaussProc}[1]{ \left\{ \xi_{#1} \right\}_{#1=1}^\infty}
\newcommand{\iidGauss}[1]{ \xi_{#1}}

\definecolor{orange}{rgb}{1,0.55,0}
\definecolor{purple}{rgb}{.7,.1,.8}
\definecolor{green}{rgb}{0,.8,0.1}

\title{\bf Bayesian Algorithms Learn to Stabilize Unknown Continuous-Time Systems}

\author{Mohamad Kazem Shirani Faradonbeh, Mohamad Sadegh Shirani Faradonbeh}
\date{}

\begin{document}

\maketitle

\begin{abstract}                
Linear dynamical systems are canonical models for learning-based control of plants with uncertain dynamics. The setting consists of a stochastic differential equation that captures the state evolution of the plant understudy, while the true dynamics matrices are unknown and need to be learned from the observed data of state trajectory. An important issue is to ensure that the system is stabilized and destabilizing control actions due to model uncertainties are precluded as soon as possible. A reliable stabilization procedure for this purpose that can effectively learn from unstable data to stabilize the system in a finite time is not currently available. In this work, we propose a novel Bayesian learning algorithm that stabilizes unknown continuous-time stochastic linear systems. The presented algorithm is flexible and exposes effective stabilization performance after a remarkably short time period of interacting with the system.
\end{abstract}



\section{Introduction}
Linear dynamical systems are ubiquitous models in the system theory for effective control of plants~\cite{chen1995linear,yong1999stochastic}. In a general setting, the state of the plant understudy follows a multidimensional stochastic differential equation depending on the state vector and control input at the time, together with the exogenous disturbance noise. An interesting problem is that of stabilization under uncertainty to ensure that the system can safely operate for a reasonable time period. 

In many situations, a reliable knowledge of the system matrices that determine the differential equations of state evolution is unavailable. Such lack of knowledges can be due to unintended internal interactions between the state quantities of an engineered plant or naturally unknown dynamics of the environment one desires to control, such as biological or financial systems~\cite{gillespie2007stochastic,lawrence2010learning,pham2009continuous}. In such situations, data-driven methods are needed to learn the underlying unknown true dynamical behaviors from the observed data. 

Accordingly, design and analysis of learning-based methods for control of unknown dynamical systems is of interest to ensure optimal system performances. For continuous-time stochastic linear systems, study of adaptive policies for near-optimal regulations has a rich literature. That consists of control policies that estimate optimal actions after long-term interactions with the system~\cite{mandl1989consistency,duncan1990adaptive,duncan1992least,duncan1999adaptive} as well as those who focus on the resulting sub-optimalities of learning-based policies~\cite{caines1992continuous,levanony2001persistent,caines2019stochastic,basei2021logarithmic,faradonbeh2021efficient}. However, algorithms for learning to stabilize with provable performance are limited to infinite time schemes for a class of controllable systems~\cite{duncan1999adaptive,caines2019stochastic}. So,  finite-time approaches are currently missing and is adopted as the subject of this work, especially for the more general case of stabilizable (rather than controlable) dynamical systems. 

Importantly, stabilization procedures require to terminate in a finite time. On one hand, the decision-maker can focus on optimality only when instabilities are fully precluded. On the other hand, during the stabilization period, the system can expose unstable behavior making the magnitude of the state extremely large. Therefore, continuation of stabilization algorithms for long periods of time defeats the purpose. For the case of discrete-times dynamics, the problem of stabilization recently received a lot of attentions. The existing results include performance guarantees for stabilizable~\cite{faradonbeh2018bfinite,faradonbeh2019randomized} and controllable systems~\cite{abbasi2011regret,lale2020explore,chen2021black}. 

In the discrete-time regime, the method of applying random feedback matrices is proposed to ensure that the resulting unknown closed-loop matrix is learnable~\cite{shirani2017non,faradonbeh2018bfinite}. Intuitively, if (the rates or directions of) explosions for an unstable linear dynamical system are indistinguishable, parameter estimates will be drastically inaccurate~\cite{faradonbeh2018finite}. To avoid these pathological cases for learning stabilizing policies, the algorithm can randomize the inputs, which in turn renders pathological closed-loop matrices unlikely~\cite{faradonbeh2018bfinite}. Then, once stabilized, low-regret reinforcement learning policies can be applied to the system to learn to minimize the operating costs. The latter trend of research includes many recent works as well, especially for the case of quadratic cost functions~\cite{faradonbeh2019applications,faradonbeh2020input,faradonbeh2020adaptive,faradonbeh2020optimism}.

In this work, we introduce an algorithm that learns to stabilize unknown continuous-time stochastic linear system in a short time period. The algorithm designs control inputs that effectively excite all modes of the system, and switches the employed feedback policies in sub-intervals of the original time interval. Then, the resulting state vectors are utilized in a Bayesian fashion to form a posterior belief about the unknown dynamics matrices. 
The performance of the algorithm is studied and effects of different parameters are discussed. We present numerical results showcasing that in a remarkably short time period, the proposed algorithm successfully stabilizes the system with high probability. Introducing a novel framework for fast stabilization, we discuss the implications and provide multiple interesting research avenues for future investigations.

This paper is organized as follows. In Section~\ref{ProblemSection}, we present the detailed statement of the problem and discuss different aspects of stabilization under uncertainty. Then, in Section~\ref{AlgoSection}, we propose a Bayesian learning-based stabilization algorithm and outline the detailed steps it relies on. Analysis of the algorithm and interpretations of the effects of different parameters are provided in Section~\ref{NumericalSection}. Finally, we spotlight the vision this work leads to, discuss possible extensions to other regimes such as large-scale and/or non-linear systems, and mention different directions for future studies on data-driven methods and algorithms for learning-to-stabilize.

\textbf{Notation:} The Euclidean norm of vectors is denoted by $\norm{\cdot}{2}$, while for matrices, we use operator norms: $\Mnorm{M}{2}= \sup\limits_{\norm{v}{2}=1} \norm{Mv}{2}$. Further, we define normal distribution \textbf{for matrices}, as follows. For $d_1 \times d_2$ matrices $M,\mu$ and $d_1 \times d_1$ matrix $\Sigma$, when we say that $M$ has the distribution $\normaldist{\mu}{\Sigma}$, it means that all $d_2$ columns of $M$ are independent from each other, and are $d_1$ dimensional normally distributed random vectors such that their covariance matrix is $\Sigma$ and their mean is the corresponding column of $\mu$. For a square matrix $M$, we denote the largest real-part of the eigenvalues of $M$ by $\mosteig{M}$. Note that $\mosteig{M}$ can be either negative (stable $M$) or non-negative (unstable $M$). Finally, $\indicator{\mathcal{S}}$ is the indicator function that has the value $1$ whenever $\mathcal{S}$ holds true, and is $0$ otherwise.

\section{Problem Statement} \label{ProblemSection}
We study stabilization of continuous-time stochastic systems under uncertainty. In this setting, the dynamics of the plant under consideration consist of the following \emph{unknown} stochastic differential equation:
\begin{equation} \label{dynamics}
\diff \state{t} = \left( \Amat{} \state{t} + \Bmat{} \action{t} \right) \diff t + \BMcoeff{t} \diff \BM{t}.
\end{equation}
That is, the open-loop transition matrix $\Amat{} \in \R^{\statedim \times \statedim}$ of the system with state $\state{t} \in \R^{\statedim}$, as well as the input matrix $\Bmat{}$ that reflects the effect of the control signal $\action{t} \in \R^{\controldim}$, both are \emph{unknown}. The differential equation describing the plant in the above model includes a white noise. That is, the system is disturbed by the Brownian motion $\BM{t}$, technically defined as a continuous-time stochastic process with independent Gaussian increments $\BM{t}-\BM{s}, s \leq t$ of mean zero, $\E{\BM{t}-\BM{s}}=0_{\noisedim}$, and covariance matrix $\cov{\BM{t}-\BM{s}}=\left(t-s\right)I_{\noisedim}$. The $\statedim \times \noisedim$ matrix $\BMcoeff{}$ that rescales the noise influence is assumed \emph{unknown} as well.

Note that if the physical behavior of the system is governed by a differential equation of a higher degree, one can convert it to the state-space model in \eqref{dynamics}. To that end, it suffices to concatenate the physical state together with its derivatives of different orders to obtain the larger state vector $\state{t}$ \cite{chen1995linear,yong1999stochastic,faradonbeh2018finite}. Therefore, studying data-driven control schemes for learning to stabilize the dynamical system in \eqref{dynamics} captures a general class of plants.

To proceed, we assume that the system is stabilizable. 
\begin{assump} \label{StabAssump}
	There exists an unknown feedback matrix $\Gainmat{\text{stab}}$ such that $\mosteig{\Amat{}+\Bmat{}\Gainmat{\text{stab}}}<0$.
\end{assump}
The above assumption is necessary. Otherwise, stability becomes infeasible even under full certainty about $\Amat{},\Bmat{}$. Moreover, thanks to linear dynamics in \eqref{dynamics}, Assumption~\ref{StabAssump} implies the mean-square stabilizability. Note that due to the Brownian noise $\BM{t}$, fluctuations and large state values in the trajectory of the stochastic system are unavoidable. However, under the stabilizing linear feedback $\action{t}=\Gainmat{\text{stab}}\state{t}$, the state $\state{t}$ becomes a multivariate Gaussian random vector with a finite covariance matrix~\cite{baldi2017stochastic}, and we have $\norm{\state{t}}{2}=\order{\sqrt{-\log \delta}}$, with probability at least $1-\delta$~\cite{faradonbeh2018finite}.

It is well-known that stabilizability is intimately related to Riccati equations. For this purpose, let the symmetric matrices $\Qmat \in \R^{\statedim \times \statedim}$ and $\Rmat \in \R^{\controldim \times \controldim}$ be positive definite. Then, for $\estA{}, M \in \R^{\statedim \times \statedim}$ and $\estB{} \in \R^{\statedim \times \controldim}$, define the mapping
\begin{equation} \label{RiccOpDeff}
\MatOpAve{\estA{},\estB{}}{M} = \estA{}^{\top} M + M \estA{} - M \estB{} \Rmat^{-1} \estB{}^{\top} M + \Qmat.
\end{equation}
So, under Assumption~\ref{StabAssump}, the algebraic continuous-time Riccati equation $\MatOpAve{\truth}{M}=0$ has a unique positive semidefinite solution $\RiccSol{\truth}$. Conversely, existence of the solution $\RiccSol{\truth}$ is sufficient for the stabilizability in Assumption~\ref{StabAssump}~\cite{chen1995linear,yong1999stochastic}. Furthermore, Riccati equations can be solved relatively fast and explicitly provide the following stabilizing linear feedback matrix~\cite{chen1995linear,yong1999stochastic}:
\begin{equation} \label{OptimalPolicy} 
\Optgain{\truth}= - \Rmat^{-1} \Bmat{}^{\top}\RiccSol{\truth}.
\end{equation}
\begin{thrm} \label{OptimalityProof}
	For the linear feedback in \eqref{OptimalPolicy}, we have
	\begin{equation*}
		\mosteig{\Amat{}+\Bmat{}\Optgain{\truth}}<0.
	\end{equation*}
\end{thrm}
However, when the exact values of the true dynamics matrices $\truth$ are not available, the Riccati equation $\MatOpAve{\truth}{M}=0$ is not solvable and the problem of stabilization is challenging. In the next section, we provide a more general framework for stabilizing the system based on approximations of $\truth$.

\subsection{Stabilization under Uncertainty}
When the true dynamics matrices are unknown, data-driven methods are required to learn from the state trajectory to stabilize the system. More precisely, learning-based algorithms first design the control signal and observe the resulting state trajectory, and then utilize the observed data to estimate $\truth$. However, learning from trajectories of limited length can never be perfectly accurate and at best can lead to approximate estimates of the dynamics matrices. Especially, since during the stabilization period the system is most likely to be exponentially unstable, this period needs to be as short as possible. In the following result, we generalize Theorem~\ref{OptimalityProof} and establish an accuracy in estimating $\truth$ that is \emph{sufficient} for stabilization.

\begin{thrm}\label{StabThm}
	Letting $\CLmat{\star}=\Amat{}+\Bmat{} \Optgain{\truth}$, assume that $\mosteig{\CLmat{\star}} \leq -1$. Then, there is a constant $\stabradii$ such that the followings are sufficient for $\mosteig{\Amat{}+ \Bmat{} \Optgain{ \estA{},\estB{} }} < 0$:
	\begin{eqnarray}
		\Mnorm{\estA{}-\Amat{}}{2} &\leq& - \stabradii \mosteig{\CLmat{\star}} \statedim^{-1/2}, \label{ThmCond1}\\
		\Mnorm{\estB{}-\Bmat{}}{arg2} &\leq& -\stabradii \mosteig{\CLmat{\star}} \statedim^{-1/2} \Mnorm{\Optgain{\truth}}{2}^{-1}. \label{ThmCond2}
	\end{eqnarray} 
\end{thrm}
The proof Theorem~\ref{StabThm} is provided in the appendix. 
Note that according to Theorem~\ref{OptimalityProof}, we have $\mosteig{\CLmat{\star}}<0$. Thus, the condition $\mosteig{\CLmat{\star}}<-1$ expresses that if one closes the loop by the linear feedback policy which is designed for $\Amat{},\Bmat{}$, then the stability margin is at least $1$. However, $\mosteig{\CLmat{\star}}<-1$ is not a fundamental requirement, is adopted merely for the ease of presentation, and extensions to the case $-1 \leq \mosteig{\CLmat{\star}} <0$ is mainly the matter of technicality.

The statement of the above result contains the following interpretations. According to Theorem~\ref{StabThm}, a coarse-grained approximation of the unknown true matrices suffices for stabilization. The accuracy specified by \eqref{ThmCond1} and \eqref{ThmCond2} indicates that systems with larger state dimension $\statedim$ are harder to stabilize, as well as those with smaller stability margins $-\mosteig{\CLmat{\star}}$. The former is caused by the larger number of parameters to learn, while the latter reflects the extent to which the system stability is robust against uncertainty. Furthermore, \eqref{ThmCond2} shows that smaller true feedback matrices $\Optgain{\Amat{},\Bmat{}}$ help stabilization because $\Mnorm{\Optgain{\Amat{},\Bmat{}}}{2}$ quantifies the strength of an effective control input for altering a possibly unstable open-loop matrix $\Amat{}$ to a stable closed-loop matrix $\CLmat{\star}$, through $\Bmat{}$. 

Note that \eqref{OptimalPolicy} does \emph{not} imply that a larger matrix $\Rmat$ leads to a smaller $\Mnorm{\Optgain{\Amat{},\Bmat{}}}{2}$ (i.e., easier stabilization). In fact, according to the Lyapunov equation
\begin{equation} \label{LyapInteg}
\RiccSol{\estA{},\estB{}} = \itointeg{0}{\infty}{ e^{\estD{}^{\top}t} \left( \Qmat+ \Optgain{\estA{},\estB{}}^{\top} \Rmat \Optgain{\estA{},\estB{}} \right) e^{\estD{}t} }{t},
\end{equation}
$\RiccSol{\Amat{},\Bmat{}}$ grows with $\Rmat$. So, by \eqref{OptimalPolicy}, it is challenging to select the best $\Qmat,\Rmat$ matrices for relaxing the accuracy in~\eqref{ThmCond2} for stabilizing the system. However, the numerical studies in the sequel suggest that smaller $\Rmat$ matrices lead to slightly larger chances for successful stabilization.

\section{Bayesian Learning Algorithm} \label{AlgoSection}
Next, we introduce a stabilization algorithm that leverages Bayesian learning methods to quickly find a stabilizing feedback for the unknown system in \eqref{dynamics}. Broadly speaking, Algorithm~\ref{algo} commits to interact with the plant in the time interval $0\leq t \leq \stablength$, and then provides a stabilizing feedback by using the collected data. It applies randomized control inputs in periods of length $\stablength/\noofepochs$, employs a Bayesian learning method to compute a posterior belief about the unknown truth, and then samples $\estpara{\stablength}$ from the posterior distribution. Finally, these sampled matrices are used together with the Riccati equation to design a stabilizing feedback. The length $\stablength$, the number of periods $\noofepochs$, and the randomization scales $\FBscale,\dithercoeff{}$, are provided to the algorithm as inputs. In the sequel, we present the detailed procedure of Algorithm~\ref{algo}.
\begin{algorithm}
	\caption{{\bf: Bayesian Learning for Stabilization} } \label{algo}
	\begin{algorithmic}
		\State{Inputs: stabilization time $\stablength$, feedback variance $\FBscale^2$, randomization variance $\dithercoeff{}^2$, number of periods $\noofepochs$}
		\State{Output: posterior samples  $\estpara{\stablength}$}\\~ 
		\State{Let $\left\{\dither{t}\right\}_{t \geq 0}$ be the randomization signal in \eqref{DitherDeff}}
		\For{period $j=1,\cdots,\noofepochs$}
		\State Sample feedback $\Gainmat{j}$ from $\normaldist{0}{\FBscale^2 I_{\statedim}}$
		\While{in the interval $\frac{\left(j-1\right) \stablength}{\noofepochs} \leq t < \frac{j \stablength}{\noofepochs}$,}
		\State Apply input $\action{t} = \Gainmat{j} \state{t}+ \dithercoeff{} \dither{t}$
		\EndWhile
		\EndFor
		\State{Calculate $\empmean{\stablength},\empiricalcovmat{\stablength}$, and $\posterior{\stablength}$ according to \eqref{RandomLSE1}, \eqref{RandomLSE2}, and \eqref{RandomLSE3}}
		\State{\textbf{Return} sample $\estpara{\stablength}$ from the posterior $\posterior{\stablength}$}
	\end{algorithmic}
\end{algorithm}

First, fix $\ditherstep>0$ and let $\left\{\dither{t}\right\}_{t \geq 0}$ be a continuous-time $\controldim$ dimensional standard Gaussian signal with independent piece-wise constant values that change after short intervals of length $\ditherstep$. Technically, suppose that $\iidGaussProc{k}$ is a sequence of independent identically distributed $\R^{\controldim}$-valued vectors that share the standard normal distribution $\normaldist{0_\controldim}{I_\controldim}$, and define the randomization signal
\begin{equation} \label{DitherDeff}
	\dither{t} = \sum\limits_{k=1}^\infty \indicator{(k-1)\ditherstep \leq t < k\ditherstep} \iidGauss{k}.
\end{equation}
That is, during the time interval $(k-1)\ditherstep \leq t < k\ditherstep$, we have $\dither{t}=\iidGauss{k}$. Note that if $\ditherstep$ shrinks, $\ditherstep^{1/2} \dither{t}$ converges in distribution to a continuous-time Gaussian process. However, in Algorithm~\ref{algo}, $\ditherstep$ is a fixed value satisfying $\ditherstep < \stablength \controldim^{-1} \noofepochs^{-1}$. Indeed, it suffices for $\ditherstep$ to be small enough to help the control signal $\action{t}$ randomizing the input for ensuring that all modes of the dynamical system in \eqref{dynamics} are properly excited. 

Algorithm~\ref{algo} breaks the interval $0 \leq t \leq \stablength$ to $\noofepochs$ sub-interval periods of equal lengths. Then, during period $j \in \left\{1,\cdots, \noofepochs\right\}$, the control input $\action{t} = \Gainmat{j} \state{t}+ \dithercoeff{} \dither{t}$ is applied to the plant, where $\Gainmat{j} \sim \normaldist{0}{\FBscale^2 I_\controldim}$ is a random feedback matrix for period $j$. Recall from the definition of normally distributed matrices in the notation that it is equivalent to say that all $\statedim$ columns of $\Gainmat{j}$ are independently drawn from a zero-mean Gaussian distribution with covariance matrix $\FBscale^2 I_\controldim$. The rationale of employing random feedbacks is to preclude the resulting true closed-loop matrix $\Amat{}+\Bmat{}\Gainmat{j}$ from falling on the pathological manifold of \emph{non-estimable} matrices~\cite{faradonbeh2018finite,faradonbeh2018bfinite,faradonbeh2019randomized}. Further technical details about necessity and effectiveness of random feedback matrices is beyond the scope of this work and can be found in the aforementioned references.

After designing and applying the above control inputs, Algorithm~\ref{algo} uses the collected observations $\left\{ \state{t},\action{t} \right\}_{0 \leq t \leq \stablength}$ to calculate a posterior distribution $\posterior{\stablength}$ on the set of $\estpara{}$ matrices, and then samples $\estpara{\stablength}$ from $\posterior{\stablength}$, as outlined below. Denote the observation at time $t$ by the vector $\statetwo{t}=\left[\state{t}^{\top},\action{t}^{\top}\right]^\top \in \R^{\statedim+\controldim}$. Then, define the $\left( \statedim+\controldim\right) \times \left( \statedim+\controldim\right)$ precision matrix 
\begin{equation} \label{RandomLSE1}
\empiricalcovmat{\stablength} = I_{\statedim+\controldim}+ \itointeg{0}{\stablength}{\statetwo{s} \statetwo{s}^{\top}}{s}, 
\end{equation}
as well as the the $\left(\statedim+\controldim\right) \times \statedim$ mean matrix
\begin{equation} \label{RandomLSE2}
\empmean{\stablength} = \empiricalcovmat{\stablength}^{-1}  \itointeg{0}{\stablength}{ \statetwo{s} }{\state{s}^{\top}}.
\end{equation}
Now, Algorithm~\ref{algo} forms the posterior belief
\begin{equation} \label{RandomLSE3}
\posterior{\stablength} = \normaldist{\empmean{\stablength}}{\empiricalcovmat{\stablength}^{-1}},
\end{equation}
about the unknown true dynamics matrices $\truth$. So, as defined in the notation, the posterior distribution of every column $i=1, \cdots, \statedim$ of $\left[\truth\right]^\top$ is a multivariate normal that its mean is the column $i$ of the matrix $\empmean{\stablength}$ and its covariance matrix is $\empiricalcovmat{\stablength}^{-1}$. Next, Algorithm~\ref{algo} samples from $\posterior{\stablength}$, and returns the resulting sample matrix $\estpara{\stablength}$. 

Finally, expecting that $\estpara{\stablength}$ is a sufficiently accurate approximation of $\truth$, the feedback matrix $\Optgain{\estA{\stablength},\estB{\stablength}}$ is expected to stabilize the system. That is, letting $\RiccSol{\estA{\stablength},\estB{\stablength}}$ be the solution of the Riccati equation $\MatOpAve{\estA{\stablength},\estB{\stablength}}{M}=0$ in \eqref{RiccOpDeff} for the dynamics parameters $\estpara{\stablength}$, we define $\Optgain{\estA{\stablength},\estB{\stablength}}=-\Rmat^{-1} \estB{\stablength}^{\top} \RiccSol{\estA{\stablength},\estB{\stablength}}$, and expect to have $\mosteig{\Amat{}+\Bmat{}\Optgain{\estA{\stablength},\estB{\stablength}}}<0$. It is straightforward to see that the probabilistic belief $\posterior{\stablength}$ is the  commonly used posterior distribution in Bayesian learning methods, if a Gaussian prior distribution about the true parameter is assumed. Of course, note that we did not assume availability of such prior beliefs about $\truth$.


\section{Performance of Learning Algorithm} \label{NumericalSection}
\begin{figure}
	\centering
	\resizebox*{9.0cm}{!}{\includegraphics{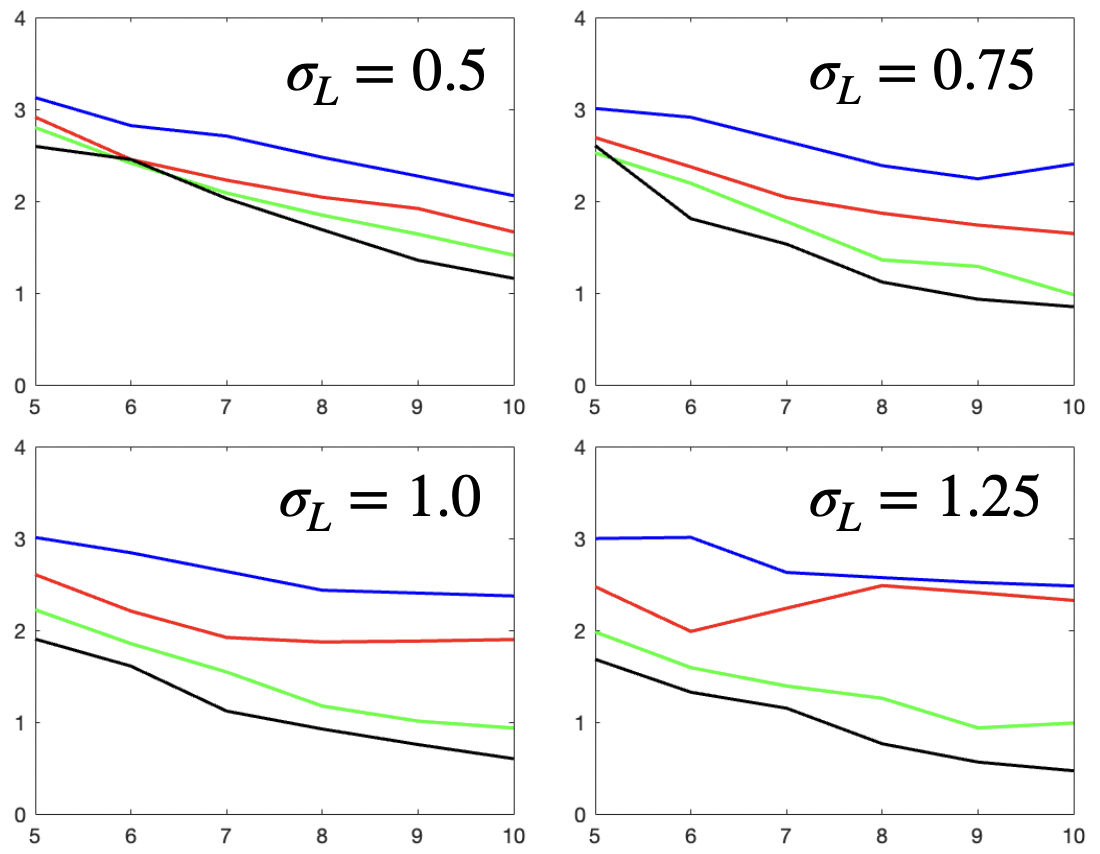}}\hspace{5pt}
	\caption{The above panel showcases the estimation error $\norm{\estpara{\stablength}-\left[\truth\right]^\top}{2}$ of Algorithm \ref{algo}, versus the stabilization time length $\stablength$. The panel consists of $4$ graphs corresponding to different values of $\FBscale$. The value of $\dithercoeff{}=1$ is fixed, while curves of different colors correspond to the number of periods (\textcolor{blue}{$\noofepochs=1$ (blue)}, \textcolor{red}{$\noofepochs = 2$ (red)}, \textcolor{green}{$\noofepochs = 3$ (green)}, and \textcolor{black}{$\noofepochs = 4$ (black)}). These plots illustrate that the algorithm is able to quickly learn the true unknown dynamical model. }
	\label{fig:error}
\end{figure}
\begin{figure}
	\centering
	\resizebox*{9.0cm}{!}{\includegraphics{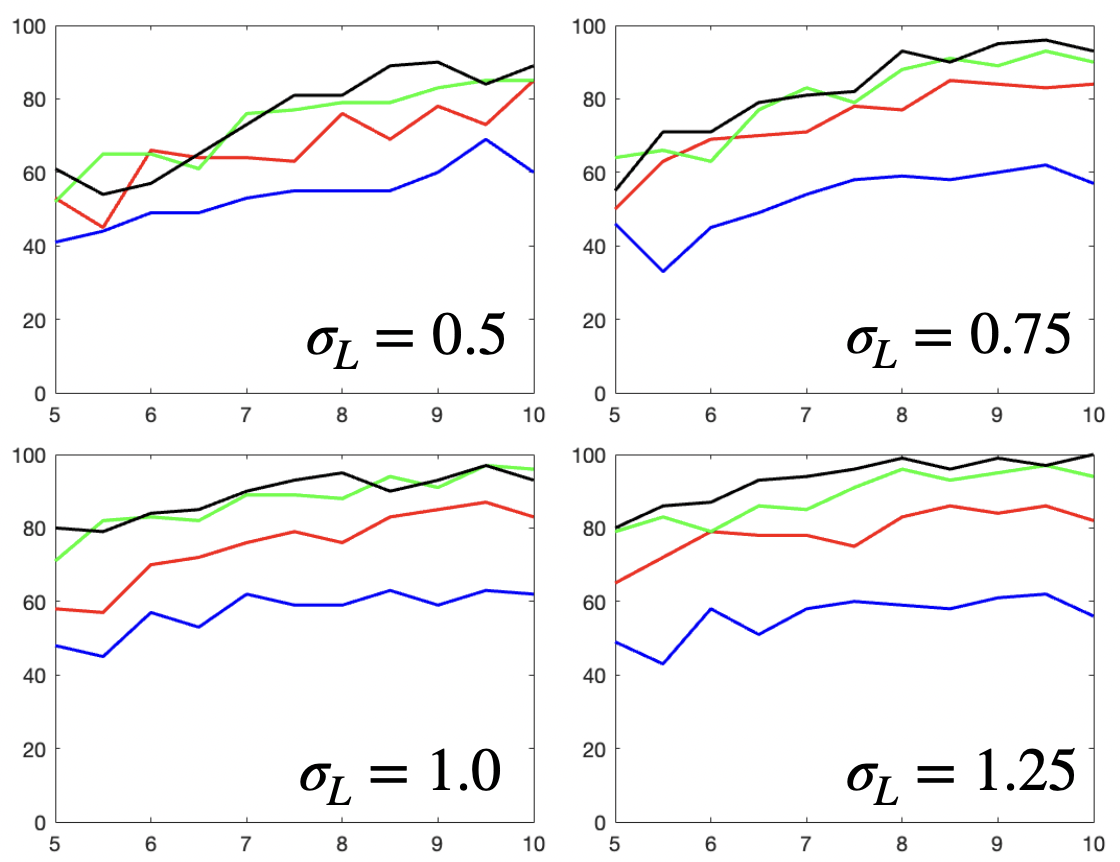}}\hspace{5pt}
	\caption{The above graphs show the percentage of successful stabilizations by Algorithm \ref{algo}, versus the stabilization time length $\stablength$. The figure consists of $4$ graphs corresponding to different values of $\FBscale$, where $\dithercoeff{}=1$. Curves of different colors represent the number of periods (\textcolor{blue}{$\noofepochs=1$ (blue)}, \textcolor{red}{$\noofepochs = 2$ (red)}, \textcolor{green}{$\noofepochs = 3$ (green)}, and \textcolor{black}{$\noofepochs = 4$ (black)}). The lower plots illustrate that with high probability, the algorithm successfully learns to stabilizes the system after the significantly short total time length of $\stablength=8$. }
	\label{fig:sigmaL}
\end{figure}
\begin{figure}
	\centering
	\resizebox*{9.0cm}{!}{\includegraphics{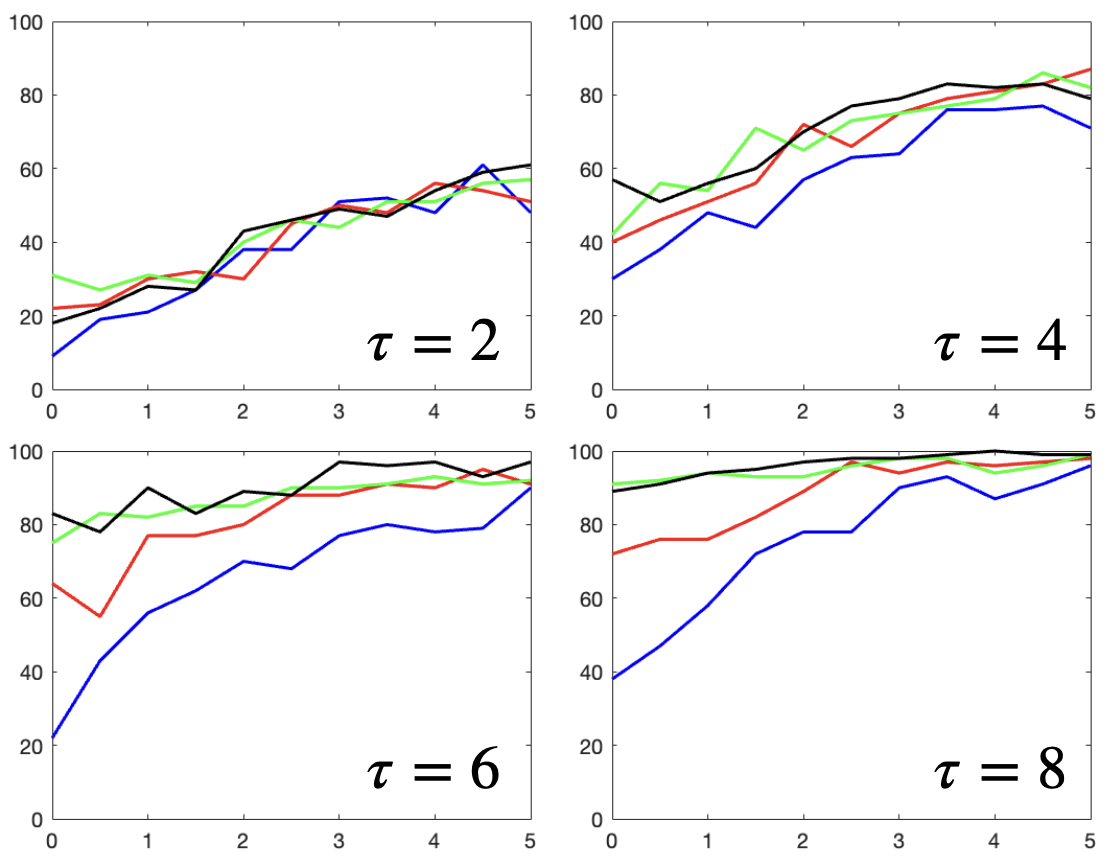}}\hspace{5pt}
	\caption{Here, we report the percentage of successful stabilizations by Algorithm \ref{algo}, versus $\dithercoeff{}$ that denotes the standard deviation of the randomization through the signal $\dither{t}$. The graphs correspond to different values of $\stablength$, while $\FBscale=1$ is fixed. Similar to before, colors distinguish the number of periods (\textcolor{blue}{$\noofepochs=1$ (blue)}, \textcolor{red}{$\noofepochs = 2$ (red)}, \textcolor{green}{$\noofepochs = 3$ (green)}, and \textcolor{black}{$\noofepochs = 4$ (black)}). This figure depicts that descent randomization in Algorithm~\ref{algo} is sufficient to stabilize the system fast.}
	\label{fig:sigmaV}
\end{figure}
\begin{figure}
	\centering
	\resizebox*{9cm}{!}{\includegraphics{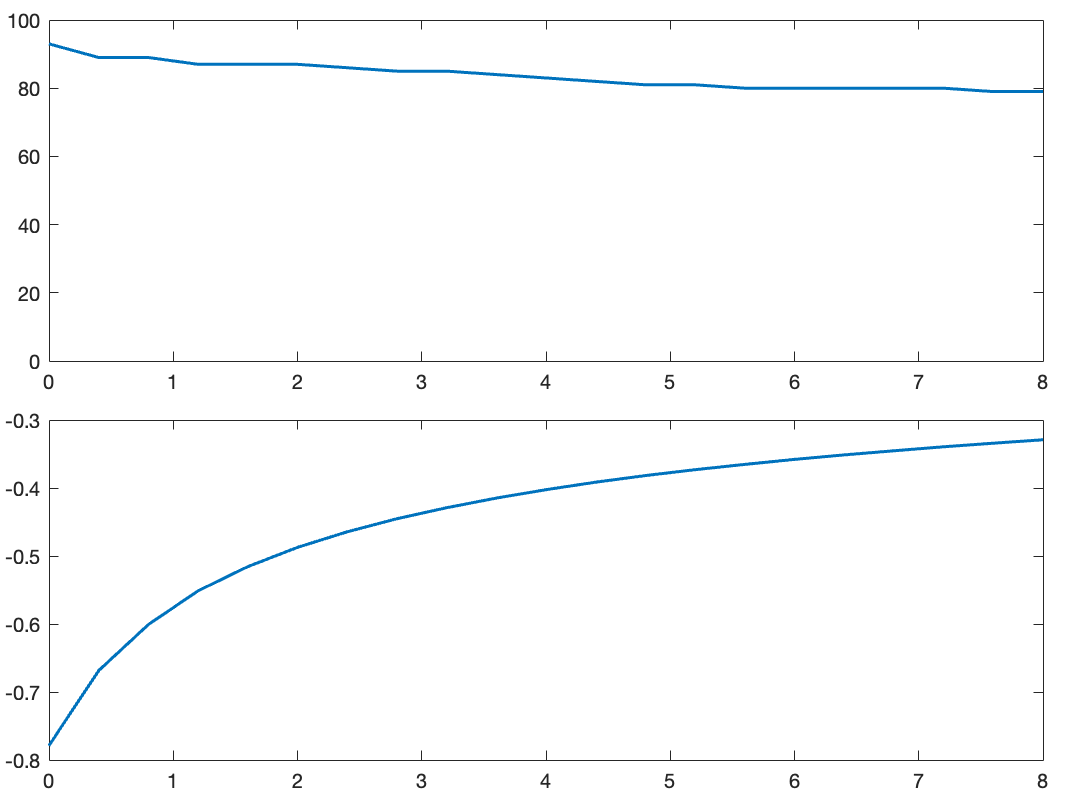}}\hspace{5pt}
	\caption{The upper graph presents the percentage of successes in learning to stabilize, versus the ratio $r$, where $\Rmat=r I_\controldim$. The lower plot is $\mosteig{\estA{\stablength}+\estB{\stablength}\Optgain{\estA{\stablength},\estB{\stablength}}}$ versus $r$. In both cases, $\FBscale=\dithercoeff{}=1, \stablength=10$. }
	\label{fig:QRratio}
\end{figure}

In this section, we analyze the performance of Algorithm~\ref{algo} by simulating a system that evolves according to \eqref{dynamics}, where the true dynamics matrices are
\begin{equation*}
	\Amat{} = \begin{bmatrix}
	&-0.46\;    &0.06\;    &0.11
	\\
	&-0.45\;    &0.27\;    &0.27
	\\
	&0.20\;    &0.18\;    &0.19
	\end{bmatrix}, \quad
	\Bmat{} = \begin{bmatrix}
	&-0.40\;    &0.44
	\\
	&0.08\;    &-0.48
	\\
	&-0.43\;    &-0.49
	\end{bmatrix}.
\end{equation*}
These matrices are chosen randomly and are relatively similar to some dynamical models for aerodynamic systems~\cite{bosworth1992linearized}. Importantly, the open-loop system is explosive; $\mosteig{\Amat{}}>0$, such that two of the three eigenvalues of the transition matrix $\Amat{}$ have positive real-parts. Therefore, stabilization of the dynamical system $\truth$ is a sufficiently challenging experiment for Algorithm~\ref{algo}.

Further, the coefficient matrix of the diffusion term in \eqref{dynamics} is  $\BMcoeff{}=I_\statedim$ in the following experiments, and the continuous-time dynamics in \eqref{dynamics} is simulated through time-discretization steps of length $10^{-3}$. Further, the value of $\ditherstep=0.2$, that reflects the length of the time pieces that the randomization signal $\dither{t}$, is fixed in the sequel. All simulations are repeated $100$ times, and the corresponding results are reported based on all replicates of the dynamical system understudy. Unless otherwise explicitly mentioned, the matrices $\Qmat,\Rmat$ in~\eqref{RiccOpDeff} are set to be identity matrices of proper dimensions.

The first set of graphical results showcase the performance of Algorithm~\ref{algo} in learning the unknown truth $\truth$, in a relatively short time period. In Fig.~\ref{fig:error}, it can be seen that as the stabilization time $\stablength$ that the algorithm spends to interact with the system and collect data increases, the accuracy of learning the unknown truth $\truth$ improves. In lights of Theorem~\ref{StabThm}, this is sufficient (though, not necessary) for fast and successful stabilization, as indicated in the following figures as well.

Fig.~\ref{fig:sigmaL} illustrates the effects of $\FBscale$ (the standard deviation of the normal distribution for the random feedback matrices $\Gainmat{j}$) in Algorithm~\ref{algo}, as well as the influence of the number of periods $\noofepochs$ and the stabilization length $\stablength$, while $\dithercoeff{}=1$ is fixed. Note that $5 \leq \stablength \leq 10$ is a remarkably short time frame for stabilization. Clearly, the proposed algorithm is able to stabilize the unknown systems very fast, especially for $\FBscale=1.0, 1.25$ that are shown in the lower two graphs. Moreover, the positive effects of $\noofepochs,\FBscale$ on the probability of successful stabilization reflect the role of \emph{diverse} random feedback matrices in both the quantitative (i.e., the number of times they are renewed; $\noofepochs$) as well as the qualitative (i.e., their magnitudes; $\FBscale$) senses. Note that $\FBscale$ cannot be too large, since otherwise, the system explodes too fast and defeats the purpose of stabilization.

In Fig.~\ref{fig:sigmaV}, we study the role of the randomization signal $\dithercoeff{}\dither{t}$ on the performance of Algorithm~\ref{algo}. For four different stabilization lengths $\stablength$, different graphs are plotted for the success rates of learning to stabilize the system versus $\dithercoeff{}$. As one can expected, larger randomizations help the algorithm to explore faster such that the probability of stabilization grows with the coefficient $\dithercoeff{}$. So, a descent randomization suffices for quickly learning to stabilize through Bayesian methods. 

The final figure of this section illustrates the effect of positive definite matrices $\Qmat,\Rmat$ in Fig.~\ref{fig:QRratio}. Recall that once Algorithm~\ref{algo} provides the sample $\estpara{\stablength}$, we let $\Qmat=I_{\statedim}$ and $\Rmat = r I_{\controldim}$, and use $\RiccSol{\estA{\stablength},\estB{\stablength}}$ that solves $\MatOpAve{\estA{\stablength},\estB{\stablength}}{M}=0$ in \eqref{RiccOpDeff}, to find $\Optgain{\estA{\stablength},\estB{\stablength}}$. Then, we apply the linear feedback policy $\action{t}=\Optgain{\estA{\stablength},\estB{\stablength}} \state{t}$ to the system to stabilize it. The upper graph in Fig.~\ref{fig:QRratio} plots the probability of stabilization versus the coefficient $r$, while the lower plot is $\mosteig{\estA{\stablength}+\estB{\stablength}\Optgain{\estA{\stablength},\estB{\stablength}}}$ when $r$ varies. 

According to these graphs, smaller $\Rmat$ matrices help stabilization. This result can be interpreted through the role of matrices $\Qmat,\Rmat$ in minimizing quadratic cost cost functions $\state{t}^\top \Qmat \state{t}+\action{t}^\top \Rmat \action{t}$, subject to the linear dynamics in~\eqref{dynamics}. It is well-known that linear feedbacks $\Optgain{\estA{},\estB{}}$ are optimal policies for that purpose~\cite{chen1995linear,yong1999stochastic}. Accordingly, for the fixed matrix $\Qmat=I_\statedim$, smaller $\Rmat$ matrices provide the control signal more freedom for moving the positive-real eigenvalues of $\Amat{}$ to the left half-plane in the complex plane. 

In addition, the second graph in Fig.~\ref{fig:QRratio} illustrates a data-driven method for optimizing the choice of $\Qmat, \Rmat$. Indeed, since in practice $\truth$ are unknown, one can optimize the stabilization procedure by selecting the matrices $\Qmat,\Rmat$ that minimize $\mosteig{\estA{\stablength}+\estB{\stablength}\Optgain{\estA{\stablength},\estB{\stablength}}}$. This corroborate the result of Theorem~\ref{StabThm} about the effect of stability margins on stabilization based on approximations of the unknown true dynamical models.

%
%
%

\section{Conclusion and Future Directions}
This work proposes the first algorithm for fast stabilization of unknown stochastic continuous-time linear dynamical systems. It is illustrated that the presented algorithm, that utilizes Bayesian methods for learning to stabilize from the observed state trajectories, performs successfully in a relatively short time period. We establish theoretical results quantifying the effects of larger dimensions and smaller stability margins on the difficulty of stabilizing an unknown system. Moreover, numerical analyses are reported showcasing the roles of different parameters on the performance, indicating that moderately larger randomizations lead to faster stabilization. 

As the first study on a canonical problem in adaptive and learning-based control, this work introduces multiple interesting problems for future work. First, establishment of theoretical performance guarantees for the proposed algorithm is of interest to understand the fundamental obstacles in learning-based stabilization policies. Besides, study of the performance in relevant problems such as network systems, non-linear dynamics, and partially observed state trajectories, are different directions for further investigations. Finally, characterizations of Bayesian learning algorithms in similar tasks in signal processing and control such as optimal output-tracking and adaptive filter design, constitute important research problems for future work.


\appendix
\section{Proof of Theorem~\ref{StabThm}}
First, define $\erterm{1}=\estA{}-\Amat{\star}+\left( \estB{}-\Bmat{\star} \right)\Optgain{\truth}$. So, \eqref{ThmCond1} and \eqref{ThmCond2} imply that 
\begin{equation} \label{Error1boundEq}
\Mnorm{\erterm{1}}{2} < -2 \stabradii \statedim^{-1/2} \mosteig{\CLmat{\star}}.
\end{equation}
Now, by applying the linear feedback matrix $\Optgain{\Amat{},\Bmat{}}$ to a linear dynamical system with dynamics matrices $\estA{},\estB{}$, we obtain the closed-loop transition matrix
\begin{equation*}
	\CLmat{1}=\estA{}+\estB{}\Optgain{\Amat{},\Bmat{}}=\CLmat{\star}+\erterm{1}.
\end{equation*}

Letting $\lambda_1, \cdots, \lambda_k$ be the eigenvalues of $\CLmat{\star}$, consider the Jordan decomposition $\CLmat{\star}=P^{-1} \Lambda P$. That is, $\Lambda$ is a block-diagonal matrix with blocks $\Lambda_1, \cdots, \Lambda_k$, where the diagonal entries of $\Lambda_i$ all are $\lambda_i$, and
\begin{eqnarray*} \label{JordanBlocks} 
\Lambda_i= \begin{bmatrix}
\lambda_i & 1 & 0 & \cdots & 0 & 0 \\
0 & \lambda_i & 1 & 0 & \cdots & 0 \\
\vdots & \vdots & \vdots & \vdots & \vdots & \vdots \\
0 & 0 & \cdots & 0 & \lambda_i & 1 \\
0 & 0 & 0 & \cdots & 0 & \lambda_i
\end{bmatrix} \in \C^{\mult{i} \times \mult{i}}.
\end{eqnarray*}

Suppose that the desired closed-loop stability does not hold. So, if $\lambda$ is an eigenvalue of $\CLmat{1}$ that its real-part is \emph{not} negative, $\mosteig{\CLmat{\star}}<0$ implies that the matrix $\CLmat{1}-\lambda I_\statedim$ is singular, while the matrix $\CLmat{\star}-\lambda I_\statedim$ is non-singular. This implies that there exists a non-zero vector $v \in \R^{\statedim}$ for which $\left(\CLmat{\star} + \erterm{1} - \lambda I_\statedim\right)v=0$. The latter leads to
\begin{equation*}
	v= - \left(\CLmat{\star} - \lambda I_\statedim\right)^{-1} \erterm{1} v = - P^{-1} \left( \Lambda - \lambda I_\statedim\right)^{-1} P \erterm{1} v.
\end{equation*}

The matrix $\Lambda-\lambda I_\statedim$ is block-diagonal, with the blocks
\begin{eqnarray*} 
	\Lambda_i - \lambda I_{\mult{i}}= \begin{bmatrix}
		\lambda_i - \lambda & 1 & 0 & \cdots & 0 & 0 \\
		0 & \lambda_i -\lambda & 1 & 0 & \cdots & 0 \\
		\vdots & \vdots & \vdots & \vdots & \vdots & \vdots \\
		0 & 0 & \cdots & 0 & \lambda_i-\lambda & 1 \\
		0 & 0 & 0 & \cdots & 0 & \lambda_i - \lambda
	\end{bmatrix} .
\end{eqnarray*}
Therefore, doing the inversion, the blocks $\left( \Lambda_i - \lambda I_{\mult{i}} \right)^{-1}$ of $\left( \Lambda - \lambda I_\statedim\right)^{-1}$ have the following forms \cite{faradonbeh2018finite,faradonbeh2021efficient}.
\begin{eqnarray*}
&& \left( \Lambda_i - \lambda I_{\mult{i}} \right)^{-1} =\\
&& - \begin{bmatrix}
\left( \lambda -\lambda_i \right)^{-1} & \left( \lambda -\lambda_i \right)^{-2}  & \cdots & \left( \lambda -\lambda_i \right)^{-\mult{i}} \\
0 & \left( \lambda -\lambda_i \right)^{-1} & \cdots & \left( \lambda -\lambda_i \right)^{-\mult{i} +1} \\
\vdots & \vdots & \vdots & \vdots \\
0 & \cdots & 0 & \left( \lambda -\lambda_i \right)^{-1}
\end{bmatrix}.
\end{eqnarray*}
Using the above formula together with the definition of operator norms for matrices, we get
\begin{equation*}
	\Mnorm{ \left(\Lambda_i - \lambda I_{\mult{i}}\right)^{-1} }{2}^2 \leq \mult{i} \left| \lambda_i - \lambda \right|^{-2},
\end{equation*}
where we used the fact that the condition $\mosteig{\CLmat{\star}} \leq -1$ in the theorem, together with the violation of stability $\Re (\lambda) \geq 0$, yields to $\left| \lambda_i - \lambda \right| \geq 1$.

Nest, putting the recent inequality on $\Mnorm{ \left(\Lambda_i - \lambda I_{\mult{i}}\right)^{-1} }{arg2}$, the block-diagonal structure of $\Lambda - \lambda I_{\statedim}$, the definition of matrix operator norms, and $v= - P^{-1} \left( \Lambda - \lambda I_\statedim\right)^{-1} P \erterm{1} v$, all together, we obtain
\begin{equation*}
	1 
	\leq \Mnorm{P^{-1}}{2} \Mnorm{P}{2} \statedim^{1/2} \left( \Re (\lambda) - \mosteig{\CLmat{\star}} \right)^{-1} \Mnorm{\erterm{1}}{2},
\end{equation*}
that is
\begin{equation*}
\Re (\lambda) \leq \mosteig{\CLmat{\star}} + \Mnorm{P^{-1}}{2} \Mnorm{P}{2} \statedim^{1/2} \Mnorm{\erterm{1}}{2}.
\end{equation*}
According to \eqref{Error1boundEq}, as long as $2 \Mnorm{P^{-1}}{2} \Mnorm{P}{2} \stabradii < 1$, it holds that $\Re (\lambda) < 0$, since Theorem~\ref{OptimalityProof} states that $\mosteig{\CLmat{\star}}<0$. 

In other words, we have 
\begin{equation} \label{AuxStability}
\mosteig{\CLmat{1}} <0,
\end{equation}
i.e., $\CLmat{1}$ is a stable closed-loop transition matrix, and we can define
\begin{equation*}
	M = \itointeg{0}{\infty}{ e^{\CLmat{1}^{\top} t} \left[ \Qmat+ \Optgain{\truth}^{\top} \Rmat \Optgain{\truth} \right] e^{\CLmat{1}t} }{t}.
\end{equation*}
Next, it is shown that applying $\Optgain{\Amat{},\Bmat{}}$ to a system of dynamics matrices $\estA{},\estB{}$, the following Lyapunov equation holds \cite{faradonbeh2021efficient}:
\begin{equation*}
M= \RiccSol{\estA{},\estB{}} + \itointeg{0}{\infty}{ e^{\CLmat{1}^{\top} t} F e^{\CLmat{1}t} }{t},
\end{equation*} 
where 
\begin{equation*}
F= \left[ \Optgain{\truth} - \Optgain{\estA{},\estB{}} \right]^{\top} \Rmat \left[ \Optgain{\truth} - \Optgain{\estA{},\estB{}} \right].
\end{equation*}
It is straightforward to see that by writing the above integral Lyapunov equation in the algebraic form, we have
\begin{eqnarray*}
&& \Qmat+ \Optgain{\truth}^{\top} \Rmat \Optgain{\truth}\\
&=&-\CLmat{1}^{\top} M - M \CLmat{1} \\
&=& -\CLmat{\star}^{\top} M - M \CLmat{\star} - \erterm{1}^{\top} M - M \erterm{1}.
\end{eqnarray*}

Leveraging stability of $\CLmat{\star}$, we obtain an integral form of the Lyapunov equation, as follows:
\begin{eqnarray*}
	& M& \\
	&=& {\small \itointeg{0}{\infty}{ e^{\CLmat{\star}^{\top} t} \left[ \Qmat+ \Optgain{\truth}^{\top} \Rmat \Optgain{\truth} + \erterm{1}^{\top} M + M \erterm{1} \right] e^{\CLmat{\star}t} }{t} }\\
	&=& \itointeg{0}{\infty}{ e^{\CLmat{\star}^{\top} t} \left[ \Qmat+ \Optgain{\truth}^{\top} \Rmat \Optgain{\truth} \right] e^{\CLmat{\star}t} }{t} \\
	&+& \itointeg{0}{\infty}{ e^{\CLmat{\star}^{\top} t} \left[ \erterm{1}^{\top} M + M \erterm{1} \right] e^{\CLmat{\star}t} }{t} \\
	&=& \RiccSol{\truth} + \itointeg{0}{\infty}{ e^{\CLmat{\star}^{\top} t} \left[ \erterm{1}^{\top} M + M \erterm{1} \right] e^{\CLmat{\star}t} }{t}.
\end{eqnarray*}

Therefore, the definition of matrix operator norm gives 
\begin{equation*}
\Mnorm{M}{2} \leq \Mnorm{\RiccSol{\truth}}{2} + 2 \Mnorm{\erterm{1}}{2} \Mnorm{M}{2} \itointeg{0}{\infty}{ \Mnorm{e^{\CLmat{\star}t}}{2}^2 }{t}.
\end{equation*}
So, according to \eqref{Error1boundEq}, if 
\begin{equation*}
-20 \stabradii \statedim^{-1/2} \mosteig{\CLmat{\star}} \itointeg{0}{\infty}{ \Mnorm{e^{\CLmat{\star}t}}{2}^2 }{t} \leq 1,
\end{equation*}
then we have
\begin{equation*}
	\Mnorm{M}{2} \leq 1.25 \Mnorm{\RiccSol{\truth}}{2}.
\end{equation*} 
On the other hand, since $F$ is a positive semidefinite matrix, the inequality $\RiccSol{\estA{},\estB{}} \leq M$ holds. Thus, we get 
\begin{equation} \label{BoundforRiccSol}
\Mnorm{\RiccSol{\estA{},\estB{}}}{2} \leq \Mnorm{M}{2} \leq 1.25 \Mnorm{\RiccSol{\truth}}{2}. 
\end{equation}

To proceed, denote the closed-loop transition matrix of the system $\estA{},\estB{}$ under its own linear feedback policy by $\estD{}=\estA{}+\estB{} \Optgain{\estA{},\estB{}}$. Further, in the Lyapunov equation in~\eqref{LyapInteg}, let $v \in \C^{\statedim}$ be such that $\norm{v}{2}=1$ and $\estD{}v = \lambda v$, where $\Re(\lambda)=\mosteig{\estD{}}$. Thus, letting $v^*$ be the transposed complex conjugate of $v$, we have
\begin{eqnarray*}
	&& v^*\RiccSol{\estA{},\estB{}}v \\
	&=& \itointeg{0}{\infty}{ v^* e^{\estD{}^{\top} t} \left[ \Qmat + \Optgain{\estA{},\estB{}}^{\top} \Rmat \Optgain{\estA{},\estB{}} \right] e^{\estD{} t}v }{t} \\
	&=& \itointeg{0}{\infty}{ \norm{\left[ \Qmat + \Optgain{\estA{},\estB{}}^{\top} \Rmat \Optgain{\estA{},\estB{}} \right]^{\frac{1}{2}} e^{\lambda t}v}{2}^2 }{t} ,
\end{eqnarray*}
So, it holds that
\begin{equation} \label{EigUpperBound}
\Mnorm{\RiccSol{\estA{},\estB{}}}{2} \geq \eigmin{\Qmat} \itointeg{0}{\infty}{ e^{2 \mosteig{\estD{}} t} }{t} \geq \frac{\eigmin{\Qmat} }{-2 \mosteig{\estD{}} }.
\end{equation}
By putting \eqref{BoundforRiccSol} and \eqref{EigUpperBound} together, we get  
\begin{equation} \label{StabilityMargin}
\mosteig{\estD{}} \leq -\eigmin{\Qmat} \left(2.5\Mnorm{\RiccSol{\truth}}{2}\right)^{-1},
\end{equation}
which provides a stability margin for the closed-loop matrix $\estD{}$.

Now, according to the above result together with \eqref{BoundforRiccSol}, we can ensure that all eigenvalues of $\Amat{}+\Bmat{}\Optgain{\estA{},\estB{}}$ are on the left open half-plane of the complex plane if $\stabradii$ is sufficiently small. To see that, observe that by $\Optgain{\estA{},\estB{}}=-\Rmat^{-1} \estB{} \RiccSol{\estA{},\estB{}}$, \eqref{BoundforRiccSol}, \eqref{ThmCond1}, and \eqref{ThmCond2}, the difference matrix
\begin{equation*}
\erterm{\star} = \Amat{}+ \Bmat{} \Optgain{\estA{},\estB{}} - \estD{}
\end{equation*} 
is arbitrarily small if $\stabradii$ is sufficiently small. Then, we can show that all eigenvalues of $\Amat{}+ \Bmat{} \Optgain{\estA{},\estB{}}$ have negative real-parts, according to \eqref{StabilityMargin}. Further details are exactly similar to the procedure of establishing \eqref{AuxStability} based on \eqref{Error1boundEq}.

~\hfill~$\blacksquare$

\end{document}